\documentclass[rnaas, modern]{aastex63}

\shorttitle{Eclipses in a candidate AE Aqr object}
\shortauthors{Littlefield \& Garnavich}

\newcommand{\lam}{LAMOST J024048.51+195226.9}

\begin{document}

\title{Identification of Orbital Eclipses in \lam, \\a Candidate AE Aqr-type Cataclysmic Variable Star}
\author{Colin Littlefield}
\affiliation{Department of Physics, University of Notre Dame, Notre Dame, IN 46556}
\affiliation{Department of Astronomy, University of Washington, Seattle, WA}
\author{Peter Garnavich}
\affiliation{Department of Physics, University of Notre Dame, Notre Dame, IN 46556}

\correspondingauthor{Colin Littlefield}
\email{clittlef@alumni.nd.edu}

%\keywords{\dots}

\begin{abstract}

    AE Aqr objects are a class of cataclysmic variable stars in which the rapidly rotating magnetosphere of the white dwarf (WD) primary centrifugally expels most infalling gas before it can accrete onto the WD. The expulsion of the accretion flow via this ``magnetic propeller'' extracts angular momentum from the WD and produces large-amplitude, aperiodic flares in optical photometry. The eponymous AE Aqr is the only confirmed member of this class of object, but recently, \citet{thorstensen} discovered a candidate AE~Aqr system: \lam. Using survey photometry, we measure a refined orbital period for this system and identify a shallow, previously unrecognized eclipse during which the system's frequent AE~Aqr-like flaring episodes cease. A dedicated follow-up study is still necessary to test the proposed AE~Aqr classification for \lam, but should it be confirmed, the eclipse of its flare-production region will offer a new means of studying the magnetic propeller phenomenon.

\end{abstract}

\section{Introduction}

AE Aqr objects are among the rarest subtypes of cataclysmic variable stars, and at present, only one has been confirmed. These short-period binaries contain a rapidly rotating, magnetized white dwarf (WD) and a late-type companion star that loses mass via Roche lobe overflow. The defining feature of this class of object is a ``magnetic propeller,'' a process in which the WD's magnetosphere centrifugally expels nearly all of the infalling matter that comes from the donor star, greatly inhibiting accretion onto the WD \citep{eh96, wynn}. The expulsion of the accretion flow extracts angular momentum from the WD's rotation, causing its spin period to gradually lengthen.

The only known AE Aqr star is AE Aqr itself, and its singular nature has motivated a voluminous body of observational and theoretical work; for reviews of AE Aqr specifically, see \citet{welsh99} and \citet{m15}. The rotational period of the WD in AE~Aqr is a mere 33~sec \citep{patterson} and gradually increasing \citep{dj94}, as expected for the magnetic-propeller scenario. AE~Aqr is famous for its erratic, large-amplitude flares, which are thought to occur when the accretion flow from the secondary is shocked---either when it encounters the WD's magnetosphere \citep{eh96} or when blobs of expelled matter collide \citep{welsh}.

\citet{thorstensen} reported a candidate AE Aqr object, \lam, with a 7.34-hour orbital period. He noted several key properties that resembled AE Aqr, including the presence of large-amplitude flares, the absence of He~II emission, and unusually weak He~I emission. Moreover, its Catalina Real-Time Transient Survey \citep[CRTS;][]{drake09} light curve (Fig.~1 in \citealt{thorstensen}) looks very similar to that of AE Aqr \citep[Fig. 6 in ][]{simon}, with one curious exception: a shallow dip between $0.9 < \phi_{orb} < 1.0$, where $\phi_{orb} = 1.0$ corresponds with the secondary's inferior conjunction (which \citealt{thorstensen} measured spectroscopically).

Here, we investigate the nature of that dip by reanalyzing the CRTS light curve, which we supplement with photometry from the All-Sky Automated Survey for Supernovae \citep[ASAS-SN;][]{shappee, kochanek}. The CRTS data offer coverage from 2005-2013, while the ASAS-SN observations commenced in 2012 and continue to this day.

\begin{figure}
    \centering
    \includegraphics[width=\textwidth]{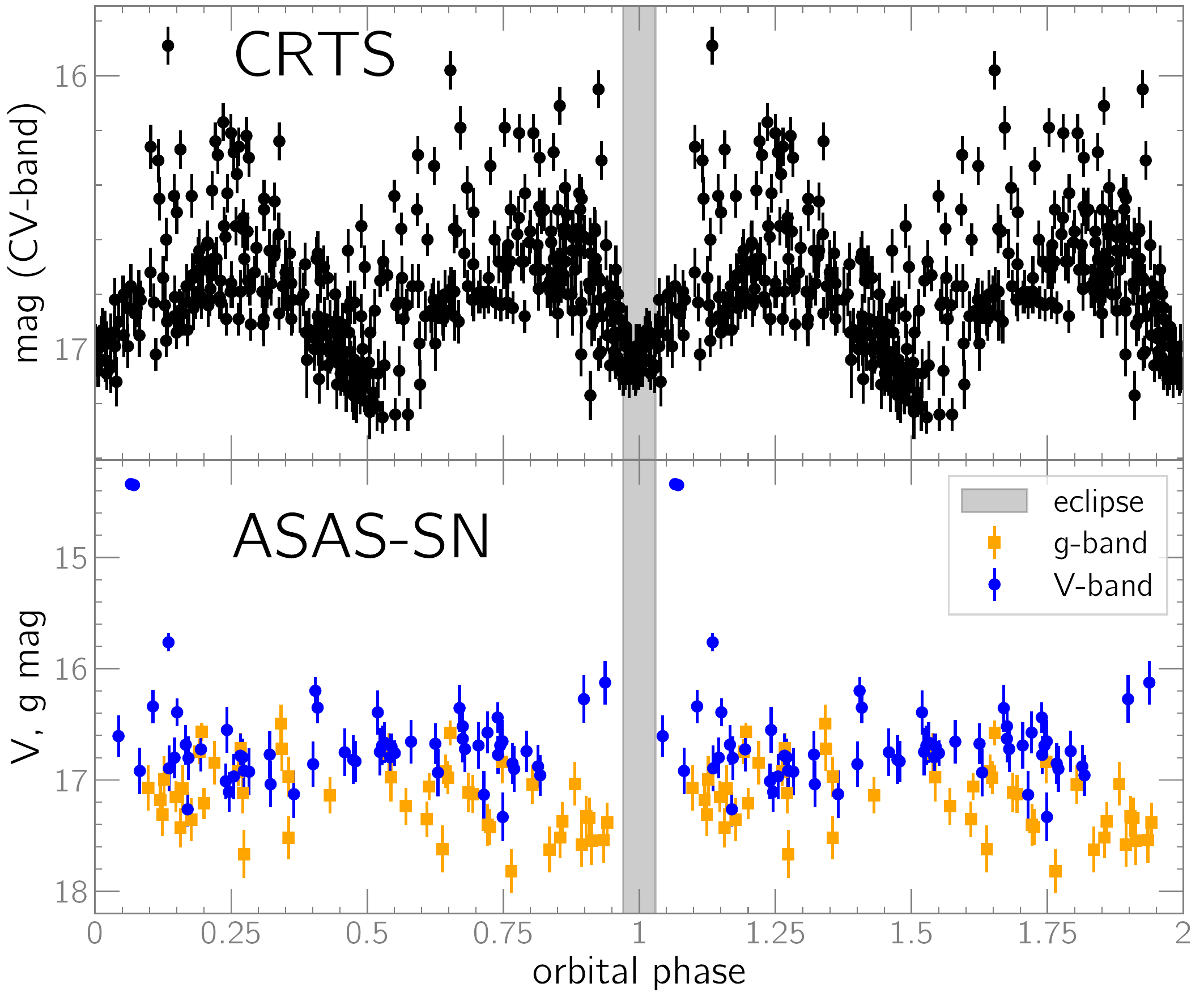}
    \caption{CRTS (top panel) and ASAS-SN (bottom panel) light curves of \lam, phased to Eq.~\ref{ephem}. The CV band for the CRTS data refers to an unfiltered bandpass using a Johnson $V$ zeropoint. ASAS-SN non-detections have been omitted for clarity, as they do not meaningfully constrain the eclipse depth. The shaded region indicates an interval in the CRTS light curve, centered on inferior conjunction, during which the high-amplitude flares cease and the low-amplitude flickering is suppressed. We interpret this as an eclipse of the flare-production site.}
    \label{fig:eclipses}
\end{figure}

\section{Analysis}

The \citet{drake14} CRTS pipeline measured a period of 0.3056840~d (no uncertainty specified), and \citet{thorstensen} adopted this value as the orbital period. However, after applying the phase-dispersion-minimization algorithm \citep{pdm} to the CRTS light curve of \lam, we refine the period to a slightly longer value of 0.3056849(5)~d; the number in parentheses is the 1$\sigma$ uncertainty on the final digit. Combining this new period with the epoch of inferior conjunction from Table~3 in \citet{thorstensen}, we obtain an updated orbital ephemeris of 
\begin{equation}
T_{conj}[BJD] = 2458836.846(2) + 0.3056849(5)\times E, \label{ephem}
\end{equation}
where $T_{conj}$ is the predicted Barycentric Julian Date (BJD) in Barycentric Dynamical Time (TDB) of inferior conjunction and $E$ is the integer cycle count.\footnote{We converted the time standard of the \citet{thorstensen} epoch of inferior conjunction from BJD in Coordinated Universal Time to BJD in TDB \citep{eastman}. }

In Fig.~\ref{fig:eclipses}, we present the CRTS and ASAS-SN light curves, phased with Eq.~\ref{ephem}. The CRTS photometry is unfiltered and therefore dominated by the contribution of the M1.5 secondary, the ellipsoidal variations of which are readily apparent in the light curve \citep{thorstensen}. With the refined period, the dip that \citet{thorstensen} observed near orbital phase 0.9 shifts in phase and becomes nearly centered on the secondary's inferior conjunction---exactly as would be expected of an eclipse of the WD by the secondary. Moreover, in the combined CRTS/ASAS-SN dataset, not a single flare is present during the dip, making this the only part of the orbit to lack them; in contrast, the CRTS light curve of AE~Aqr \citep[Fig. 6 in ][]{simon} shows flares at all orbital phases. Based on these arguments, we interpret the dip as an eclipse by the secondary. Although the $\sim$0.2-mag depth of the eclipse might seem unusually shallow, the red-sensitive CRTS bandpass emphasizes the contribution of the secondary, thereby reducing the fractional contribution of the WD and diluting the eclipse depth. In Sec.~\ref{sec:discussion}, we discuss how additional spectroscopy and photometry in \citet{thorstensen} support the eclipse interpretation. 

The ASAS-SN light curve in Fig.~\ref{fig:eclipses} is quite sparse and does not contain any detections during the eclipse. Both properties result from \lam\ being close to ASAS-SN's limiting magnitude. As a result, even a shallow eclipse could cause the system to fade below the survey's detection threshold.

Of the observational data analyzed by \citet{thorstensen}, only the CRTS light curve has a sufficiently long baseline to be significantly impacted by our revised ephemeris. The CRTS data were obtained between 2005 and 2013, whereas the epoch of inferior conjunction was measured in 2019.96. In contrast, the \citet{thorstensen} time-series photometry and spectroscopy (his Figs.~3~and~4) were obtained within $\sim$1~month of the epoch of inferior conjunction, which is too short of a baseline to notice the effects of the refined orbital period.

\section{Discussion \& Conclusion}\label{sec:discussion}

The identification of an eclipse at inferior conjunction helps to explain several interesting features in the data presented by \citet{thorstensen}. His Fig.~3 shows that He~I $\lambda$6678~\AA\ emission briefly disappears at inferior conjunction, as does much of the H$\alpha$ line (particularly its high-velocity wings). This behavior can be easily explained by an eclipse of the corresponding line-forming regions. Likewise, Fig.~4 in \citet{thorstensen} shows that in several different photometric time series, a shallow dip occurs at inferior conjunction. The morphology of that dip experiences significant orbit-to-orbit variation, suggesting that the eclipsed structure changes appreciably on orbital timescales.

If we assume, for the sake of argument, that \lam\ is a member of the AE~Aqr club, the eclipses will provide new insight into the magnetic propeller phenomenon. For example, the eclipse of the flare-production region implies that it is relatively close to the WD, likely favoring models in which flares occur when blobs of infalling matter are shocked as they encounter the magnetosphere \citep{eh96}, as opposed to scenarios in which collisions between expelled blobs at large distances from the WD are the culprit \citep{welsh}. Although these prospects are enticing, follow-up observations are still necessary to confirm that \lam\ is an AE~Aqr object, and the detection of a very short spin period will be a particularly critical test of that hypothesis \citep{thorstensen}.

\acknowledgments

We thank John Thorstensen and Paula Szkody for helpful discussions.

\bibliography{bib.bib}

\begin{thebibliography}{}
\expandafter\ifx\csname natexlab\endcsname\relax\def\natexlab#1{#1}\fi
\providecommand{\url}[1]{\href{#1}{#1}}
\providecommand{\dodoi}[1]{doi:~\href{http://doi.org/#1}{\nolinkurl{#1}}}
\providecommand{\doeprint}[1]{\href{http://ascl.net/#1}{\nolinkurl{http://ascl.net/#1}}}
\providecommand{\doarXiv}[1]{\href{https://arxiv.org/abs/#1}{\nolinkurl{https://arxiv.org/abs/#1}}}

\bibitem[{{de Jager} {et~al.}(1994){de Jager}, {Meintjes}, {O'Donoghue}, \&
  {Robinson}}]{dj94}
{de Jager}, O.~C., {Meintjes}, P.~J., {O'Donoghue}, D., \& {Robinson}, E.~L.
  1994, \mnras, 267, 577, \dodoi{10.1093/mnras/267.3.577}

\bibitem[{{Drake} {et~al.}(2009){Drake}, {Djorgovski}, {Mahabal}, {Beshore},
  {Larson}, {Graham}, {Williams}, {Christensen}, {Catelan}, {Boattini},
  {Gibbs}, {Hill}, \& {Kowalski}}]{drake09}
{Drake}, A.~J., {Djorgovski}, S.~G., {Mahabal}, A., {et~al.} 2009, \apj, 696,
  870, \dodoi{10.1088/0004-637X/696/1/870}

\bibitem[{{Drake} {et~al.}(2014){Drake}, {Graham}, {Djorgovski}, {Catelan},
  {Mahabal}, {Torrealba}, {Garc{\'\i}a-{\'A}lvarez}, {Donalek}, {Prieto},
  {Williams}, {Larson}, {Christen sen}, {Belokurov}, {Koposov}, {Beshore},
  {Boattini}, {Gibbs}, {Hill}, {Kowalski}, {Johnson}, \& {Shelly}}]{drake14}
{Drake}, A.~J., {Graham}, M.~J., {Djorgovski}, S.~G., {et~al.} 2014, \apjs,
  213, 9, \dodoi{10.1088/0067-0049/213/1/9}

\bibitem[{{Eastman} {et~al.}(2010){Eastman}, {Siverd}, \& {Gaudi}}]{eastman}
{Eastman}, J., {Siverd}, R., \& {Gaudi}, B.~S. 2010, \pasp, 122, 935,
  \dodoi{10.1086/655938}

\bibitem[{{Eracleous} \& {Horne}(1996)}]{eh96}
{Eracleous}, M., \& {Horne}, K. 1996, \apj, 471, 427, \dodoi{10.1086/177979}

\bibitem[{{Kochanek} {et~al.}(2017){Kochanek}, {Shappee}, {Stanek}, {Holoien},
  {Thompson}, {Prieto}, {Dong}, {Shields}, {Will}, {Britt}, {Perzanowski}, \&
  {Pojma{\'n}ski}}]{kochanek}
{Kochanek}, C.~S., {Shappee}, B.~J., {Stanek}, K.~Z., {et~al.} 2017, \pasp,
  129, 104502, \dodoi{10.1088/1538-3873/aa80d9}

\bibitem[{{Meintjes} {et~al.}(2015){Meintjes}, {Odendaal}, \& {van
  Heerden}}]{m15}
{Meintjes}, P.~J., {Odendaal}, A., \& {van Heerden}, H. 2015, Acta Polytechnica
  CTU Proceedings, 2, 86

\bibitem[{{Patterson}(1979)}]{patterson}
{Patterson}, J. 1979, \apj, 234, 978, \dodoi{10.1086/157582}

\bibitem[{{Shappee} {et~al.}(2014){Shappee}, {Prieto}, {Grupe}, {Kochanek},
  {Stanek}, {De Rosa}, {Mathur}, {Zu}, {Peterson}, {Pogge}, {Komossa}, {Im},
  {Jencson}, {Holoien}, {Basu}, {Beacom}, {Szczygie{\l}}, {Brimacombe},
  {Adams}, {Campillay}, {Choi}, {Contreras}, {Dietrich}, {Dubberley},
  {Elphick}, {Foale}, {Giustini}, {Gonzalez}, {Hawkins}, {Howell}, {Hsiao},
  {Koss}, {Leighly}, {Morrell}, {Mudd}, {Mullins}, {Nugent}, {Parrent},
  {Phillips}, {Pojmanski}, {Rosing}, {Ross}, {Sand}, {Terndrup}, {Valenti},
  {Walker}, \& {Yoon}}]{shappee}
{Shappee}, B.~J., {Prieto}, J.~L., {Grupe}, D., {et~al.} 2014, \apj, 788, 48,
  \dodoi{10.1088/0004-637X/788/1/48}

\bibitem[{{Stellingwerf}(1978)}]{pdm}
{Stellingwerf}, R.~F. 1978, \apj, 224, 953, \dodoi{10.1086/156444}

\bibitem[{{Thorstensen}(2020)}]{thorstensen}
{Thorstensen}, J.~R. 2020, \aj, 160, 151, \dodoi{10.3847/1538-3881/aba7c7}

\bibitem[{{{\v{S}}imon}(2020)}]{simon}
{{\v{S}}imon}, V. 2020, \pasj, 72, 35, \dodoi{10.1093/pasj/psaa012}

\bibitem[{{Welsh}(1999)}]{welsh99}
{Welsh}, W.~F. 1999, in Astronomical Society of the Pacific Conference Series,
  Vol. 157, Annapolis Workshop on Magnetic Cataclysmic Variables, ed.
  C.~{Hellier} \& K.~{Mukai}, 357

\bibitem[{{Welsh} {et~al.}(1998){Welsh}, {Horne}, \& {Gomer}}]{welsh}
{Welsh}, W.~F., {Horne}, K., \& {Gomer}, R. 1998, \mnras, 298, 285,
  \dodoi{10.1046/j.1365-8711.1998.01643.x}

\bibitem[{{Wynn} {et~al.}(1997){Wynn}, {King}, \& {Horne}}]{wynn}
{Wynn}, G.~A., {King}, A.~R., \& {Horne}, K. 1997, \mnras, 286, 436,
  \dodoi{10.1093/mnras/286.2.436}

\end{thebibliography}

\end{document}